\documentclass[11pt,a4paper]{article}
\pdfoutput=1
\usepackage{jheppub}

\usepackage{etoolbox}% http://ctan.org/pkg/etoolbox
    
 \patchcmd{\maketitle}{\@fpheader}{}{}{}

\usepackage{bbding}
\usepackage{pifont}
\usepackage{wasysym}
\usepackage{wrapfig}
\usepackage{amsmath}
\usepackage{verbatim}
\usepackage{amssymb}
\usepackage{inputenc}
\usepackage{textcomp}
\usepackage{appendix}
\usepackage{mathrsfs}
\usepackage{float}
\usepackage{prettyref}
\usepackage{xfrac}

\restylefloat{table}

\newcommand{\be}[1]{\begin{equation}\label{#1} }
\newcommand{\ee}{\end{equation}}
\newcommand{\bea}[1]{\begin{eqnarray}\label{#1} }
\newcommand{\eea}{\end{eqnarray}}

\newcommand{\bes}{\begin{subequations}}
\newcommand{\ees}{\end{subequations}}

\title{\boldmath  BMS$_{3}$ (Carrollian) field theories from a bound in the coupling of current-current deformations of CFT$_{2}$}

\author[a,b]{Pulastya Parekh,}
\author[c]{David Tempo,}
\author[a,b]{Ricardo Troncoso}
\affiliation[a]{Centro de Estudios Cient\'{i}ficos (CECs), Av. Arturo Prat 514, Valdivia, Chile.}
\affiliation[b]{Facultad de Ingenier\'{i}a, Arquitectura y Dise\~{n}o, Universidad San Sebasti\'{a}n, sede Valdivia,
General Lagos 1163, Valdivia, Chile.}
\affiliation[c]{Departamento  de  Ciencias  Matem\'{a}ticas  y  F\'{i}sicas, Universidad  Cat\'{o}lica  de  Temuco,  Montt  56,  Casilla  15-D,  Temuco,  Chile.}
\emailAdd{parekh@cecs.cl}
\emailAdd{jtempo@uct.cl}
\emailAdd{ricardo.troncoso@uss.cl}

\abstract{Two types of Carrollian field theories are shown to emerge from finite
current-current deformations of toroidal CFT$_{2}$'s when the deformation
coupling is precisely fixed, up to a sign. In both cases the energy and momentum densities fulfill the BMS$_{3}$ algebra.
Applying these results to the bosonic string, one finds that the electric-like deformation (positive coupling) reduces to the standard tensionless
string. The magnetic-like deformation (negative coupling) yields to a new theory, still being relativistic, devoid of tension and endowed with an ``inner
Carrollian structure''. Classical solutions describe a sort of ``self-interacting
null particle'' moving along generic null curves of the original
background metric, not necessarily geodesics. This magnetic-like theory
is also shown to be recovered from inequivalent limits in the tension
of the bosonic string. Electric- and magnetic-like deformations of toroidal CFT$_{2}$'s
can be seen to correspond to limiting cases of continuous exactly
marginal (trivial) deformations spanned by an SO(1,1) automorphism
of the current algebra. Thus, the absolute value of the current-current
deformation coupling is shown to be bounded. When the bound saturates, the deformation ceases
to be exactly marginal, but still retains the full conformal symmetry in two alternative
ultrarelativistic regimes. 
}

\makeatother

\begin{document}
\maketitle \flushbottom \newpage{}

\section{Introduction}

Continuous current-current deformations have played an important role
in order to unveil the structure of the moduli spaces of two-dimensional
CFTs \cite{Dijkgraaf:1987vp,Chaudhuri:1988qb,Hassan:1992gi,Kiritsis:1993ju,Henningson:1992rn,Forste:2003km}.
In particular, for CFT's endowed with left and right Kac-Moody symmetries,
deformations by marginal operators of the form
\begin{equation}
g\;c_{IK}\;J^{I}\bar{J}^{K}\;,
\end{equation}
have been shown to be exactly (or integrably) marginal provided that
the combination is constructed from the subset of Abelian currents
$J^{I}$, $\bar{J}^{K}$, belonging to the Cartan subalgebra, being
preserved under the class of deformations. As pointed out in \cite{Chaudhuri:1988qb},
although it is implicitly assumed that the coupling $g$ is a perturbation,
the result was argued to be valid to all orders in the deformation
parameter, and even holds for finite values of $g$ in certain classes
of CFTs \cite{Hassan:1992gi,Kiritsis:1993ju,Henningson:1992rn,Forste:2003km}. 

One might then wonder whether the deformations remain valid for arbitrarily
large values of the deformation parameter, or if the moduli spaces
might acquire a boundary in case of a bound in the coupling. In this
sense, it has been recently shown that $g$ turns out to be bounded
even in the simple case of a single free boson \cite{Rodriguez:2021tcz,Tempo:2022ndz},
in which the current-current deformation overlaps with the so-called
$\sqrt{T\bar{T}}$ one\footnote{These generically marginal deformations have also been studied in
\cite{Conti:2022egv,Ferko:2022cix,Babaei-Aghbolagh:2022leo,Hou:2022csf,Ferko:2023sps,Ebert:2023tih,Ferko:2023ozb,Garcia:2022wad}
and they are manifestly different from the well-known irrelevant $T\bar{T}$
ones \cite{Zamolodchikov:2004ce,Smirnov:2016lqw,Cavaglia:2016oda},
whose properties are explored in e.g., \cite{McGough:2016lol,Aharony:2018bad,Gorbenko:2018oov,Conti:2019dxg,Guica:2019nzm,Jorjadze:2020ili,He:2022zcf}
(for a review also see \cite{Jiang:2019epa}).}. The latter class of deformations turns out to be generically marginal
and nontrivial, characterized by a deformation coupling whose absolute
value is bounded. When the bound saturates, the deformation ceases
to be exactly marginal since it yields to two classes of field theories
with conformal Carrollian (BMS$_{3}$) symmetries. The BMS$_{3}$
generators arise from a nonlinear map of the generators of the conformal
algebra in 2D, so that the momentum density remains the same, $P=\bar{T}-T$,
while the energy density (generator of supertranslations) is of the
form
\begin{equation}
{\cal H}_{(\pm)}=T+\bar{T}\pm2\sqrt{T\bar{T}}\;.\label{eq:Hmn(BMS3)}
\end{equation}
Thus, both sets of generators $\{{\cal H}_{(+)},P\}$ and $\{{\cal H}_{(-)},P\}$
fulfill the BMS$_{3}$ algebra, and yield to different theories of
``electric and magnetic type'', respectively. It is worth highlighting
that no limiting process is involved in the mapping \eqref{eq:Hmn(BMS3)}
that relates BMS$_{3}$ and the conformal algebra generators. Furthermore,
as pointed out in \cite{Tempo:2022ndz}, both sets are related through
a nonlinear automorphism of BMS$_{3}$, defined as 
\begin{equation}
{\cal H}_{(+)}{\cal H}_{(-)}=P^{2}\;,\label{eq:BMS3-Auto}
\end{equation}
so that theories of electric and magnetic type turn out to be dual
to each other in this sense. 

The BMS$_{3}$ generators ${\cal H}_{(\pm)}$ can also be seen to
emerge from a limiting case of the continuous nonlinear automorphism
of the conformal algebra in 2D, parameterized by $\alpha$, defined
as \cite{Tempo:2022ndz}\footnote{Here we follow a different sign convention as compared with \cite{Tempo:2022ndz},
i.e.: $\alpha$$_{\text{here}}=-\alpha_{\text{there}}$.}

\begin{equation}
{\cal H}_{(\alpha)}=\cosh\left(\alpha\right)\left(T+\bar{T}\right)+2\sinh\left(\alpha\right)\sqrt{T\bar{T}}\;,
\end{equation}
so that the momentum density $P$ and the spectrally flowed energy
density ${\cal H}_{(\alpha)}$ preserve the conformal algebra. Indeed,
rescaling the energy density according to

\begin{equation}
\tilde{{\cal H}}_{(\alpha)}:=\frac{{\cal H}_{(\alpha)}}{\cosh\left(\alpha\right)}=T+\bar{T}+2\tanh\left(\alpha\right)\sqrt{T\bar{T}}\;,
\end{equation}
the supertranslation generators of the BMS$_{3}$ algebra \prettyref{eq:Hmn(BMS3)}
are recovered when $\alpha\rightarrow\pm\infty$, i.e., $\tilde{{\cal H}}_{(\pm\infty)}={\cal H}_{(\pm)}$,
which corresponds to finite deformations with coupling $g=\pm2$.

In sum, the continuous $\sqrt{T\bar{T}}$ deformation of the Hamiltonian
action 

\begin{equation}
I_{(\alpha)}=I_{(0)}+2\tanh\left(\alpha\right)\int d\tilde{t}d\phi\sqrt{T\bar{T}}\;,
\end{equation}
generically preserves the conformal symmetry, unless $\alpha\rightarrow\pm\infty$
($g=\pm2$) since in that cases the conformal symmetries deform to
conformal Carrollian ones (BMS$_{3}$).

Therefore, a bound in the coupling of current-current deformations
also arises when they overlap with the $\sqrt{T\bar{T}}$ ones, as
it is the case of a single free boson
\begin{equation}
I_{(0)}\left[\varphi\right]=-\frac{1}{2}\int d^{2}x\sqrt{-g}\partial_{\mu}\varphi\partial^{\mu}\varphi\ .
\end{equation}
Indeed, in this case $T=J^{2}$ and $\bar{T}=\bar{J}^{2}$, and hence
\begin{equation}
\sqrt{T\bar{T}}=\pm J\bar{J}\;.
\end{equation}
Thus, the continuous $\sqrt{T\bar{T}}$ deformation of a single free
boson is exactly marginal, and actually trivial because it just amounts
to a rescaling of the action ($I_{(\alpha)}=e^{\alpha}I_{(0)}$),
as it has been recently shown in \cite{Conti:2022egv,Ferko:2022cix,Babaei-Aghbolagh:2022leo}
(see also \cite{Hou:2022csf}) from a Lagrangian approach, and in
\cite{Tempo:2022ndz} in the Hamiltonian one. Nonetheless, according
to \cite{Tempo:2022ndz}, it follows that the coupling of the current-current
deformation of a single free boson is bounded as $\left|g\right|\leq2$,
so that the deformation is trivial when the bound is strictly fulfilled,
while it becomes ultra-relativistic when it saturates ($g=\pm2$),
yielding to two inequivalent BMS$_{3}$ field theories described by
\cite{Rodriguez:2021tcz}
\begin{equation}
I_{(+)}=\int d^{2}x\left(\pi\dot{\varphi}-\pi^{2}\right)\ \ ,\ \ I_{(-)}=\int d^{2}x\left(\pi\dot{\varphi}-\varphi^{\prime2}\right)\ .
\end{equation}
Both actions $I_{(+)}$ and $I_{(-)}$ were then shown to be alternatively
recovered from different Carrollian limits of electric and magnetic
type, respectively when the speed of light $c\rightarrow0$ \cite{Henneaux:2021yzg}
(see also \cite{Duval:2014uoa})\footnote{The Carrollian limit of electric type is taken in a similar way as
in the case of the tensionless string \cite{Gamboa:1990wh,Lindstrom:1990qb,Isberg:1992ia,Isberg:1993av,Bagchi:2013bga,Bagchi:2015nca,Bagchi:2016yyf,Bagchi:2020fpr}.
The quantization of the theory described by $I_{(+)}$ has been addressed
in \cite{Hao:2021urq} (see also \cite{Saha:2022gjw,Banerjee:2023jpi}), and further
aspects have been explored in \cite{Chen:2022cpx,Bagchi:2022eav,Banerjee:2022ime}.
In higher dimensions, the action $I_{(+)}$ has been discussed
in \cite{Bergshoeff:2017btm} and more recently in e.g. \cite{Figueroa-OFarrill:2023vbj}.}. Additionally they can be seen to arise from switching off the self-interaction
of the \textquotedblleft flat versions'' of Liouville theory in \cite{Barnich:2012aw},
that were obtained through similar \textquotedblleft flat limits''
of standard Liouville theory. 

One of the main aims of our work is to extend these results for current-current
deformations of the single free boson to the case of toroidal CFT$_{2}$
endowed with $N+N$ Abelian currents. Some work along these lines
has already been done in \cite{Bagchi:2022nvj} through a class of
\textquotedblleft infinite boosts'' spanned by certain degenerate
(non-invertible) linear transformations acting on the coordinates,
that agrees with some of our results of the electric-like deformation
discussed in section \ref{sec:Electric-like deformation}. 

The plan of the paper is organized as follows. Section \ref{sec:BMS-field-theories}
is devoted to show how BMS$_{3}$ field theories emerge from two classes
of finite current-current deformations of toroidal CFT$_{2}$. These
inequivalent deformations are dubbed electric- and magnetic-like
ones. In section \ref{sec:deformationsBosonicString-III} we apply
these deformations to the bosonic string, where we find that the electric-like
one agrees with the standard tensionless limit; while in the magnetic-like
case we obtain a new theory that classically behaves as a sort of
\textquotedblleft self-interacting null particle\textquotedblright{}
that can also be obtained from different limits in the tension. In
section \ref{sec:BMS-deformationsIV} we explain how the electric-
and magnetic-like BMS$_{3}$ deformations correspond to limiting cases
of continuous exactly marginal deformations spanned by an $SO(1,1)$
automorphism of the algebra of left and right currents. In particular,
the absolute value of the current-current deformation coupling turns
out to be bounded, so that both types of BMS$_{3}$ field theories
arise when the bound saturates. In other words, when the bound saturates
the deformation ceases to be exactly marginal, but remains being fully
conformal in two different ultrarelativistic regimes. We conclude
in section \ref{sec:EndingRemarks-V} with some ending remarks.

\section{BMS$_{3}$ field theories emerging from finite $J\bar{J}$ deformations
of CFT$_{2}$\label{sec:BMS-field-theories}}

Let us consider a class of classical bosonic toroidal CFT$_{2}$,
whose left and right chiral (holomorphic) components of the stress
energy tensor can be written in terms of $N$ left and right Abelian
currents, according to 
\begin{equation}
T=g_{IK}\,J^{I}J^{K}\;\;,\;\;\bar{T}=g_{IK}\,\bar{J}^{I}\bar{J}^{K}\;,
\end{equation}
where $g_{IJ}$ stands for the inverse of the invariant bilinear form
in the algebra of the currents, whose Poisson brackets are given by
\begin{align}
\left\{ J^{I}(\phi),J^{K}(\varphi)\right\} =-g^{IK}\partial_{\phi}\delta(\phi-\varphi)\;\;,\;\; & \left\{ \bar{J}^{I}(\phi),\bar{J}^{K}(\varphi)\right\} =g^{IK}\partial_{\phi}\delta(\phi-\varphi)\;.\label{eq:currentsPoissonB}
\end{align}
Expanding in modes, the conformal algebra generators\footnote{In our conventions the mode expansions read $J^{I}\left(\phi\right)=\frac{1}{2}\sum_{n}\bar{J}_{n}^{I}e^{-in\phi}$,
$\bar{J}^{I}\left(\phi\right)=\frac{1}{2}\sum_{n}J_{n}^{I}e^{in\phi}$,
as well as $T\left(\phi\right)=\frac{1}{2}\sum_{n}\bar{L}_{n}e^{-in\phi}$,
$\bar{T}\left(\phi\right)=\frac{1}{2}\sum_{n}L_{n}e^{in\phi}$. Here
and afterwards we use $A\cdot B:=g_{IJ}A^{I}B^{J}$ and $A^{2}:=g_{IJ}A^{I}A^{J}$.} 
\begin{equation}
L_{n}=\frac{1}{2}\sum_{m}J_{n-m}\cdot J_{m}\;\;,\;\;\bar{L}_{n}=\frac{1}{2}\sum_{m}\bar{J}_{n-m}\cdot\bar{J}_{m}\;,
\end{equation}
with the Abelian currents fulfill two copies of the semidirect sum
of de Witt algebra with a chiral Abelian current algebra, i.e.,
\begin{align}
[L_{m},L_{n}] & =(m-n)L_{m+n}\;\;,\;\;[L_{n},J_{m}^{I}]=-mJ_{m+n}^{I}\;,
\end{align}
\begin{equation}
[J_{n}^{I},J_{m}^{K}]=ng^{IK}\delta_{m+n,0}\;,
\end{equation}
where $\left[\cdot,\cdot\right]=i\left\{ \cdot,\cdot\right\} $, and
similarly for $\bar{L}_{m}$ and $\bar{J}_{m}^{I}$.

For our purposes, it is useful to change the basis so that the generators
of the conformal symmetry correspond to the energy and momentum densities,
respectively given by
\begin{align}
H & =\bar{T}+T\;\;,\;\;P=\bar{T}-T\;,\label{eq:H,P}
\end{align}
while the currents are redefined as
\begin{equation}
k_{(\pm)}^{I}=\bar{J}^{I}\pm J^{I}\;,\label{eq:kpm}
\end{equation}
so that under parity, $k_{(+)}^{I}$ becomes even and $k_{(-)}^{I}$
is odd. Note that the energy and momentum densities in terms of the
currents read
\begin{align}
H & =\bar{J}^{2}+J^{2}=\frac{1}{2}\left(k_{(+)}^{2}+k_{(-)}^{2}\right)\;,\label{eq:Hjjbkmkn}\\
P & =\bar{J}^{2}-J^{2}=k_{(+)}\cdot k_{(-)}\;.\label{eq:Pjjbkmkn}
\end{align}
The generators of the full algebra in the new basis are then given
by the set $\{H,P,k_{(+)},k_{(-)}\}$. Expanding in modes according
to $X(\phi)=\frac{1}{2}\sum_{n}X_{n}e^{in\phi}$, equations \eqref{eq:H,P}
and \eqref{eq:kpm} then read
\begin{align}
H_{n} & =L_{n}+\bar{L}_{-n}\;\;,\;\;P_{n}=L_{n}-\bar{L}_{-n}\;,
\end{align}
\begin{equation}
k_{(\pm)n}^{I}=J_{n}^{I}\pm\bar{J}_{-n}^{I}\;,
\end{equation}
while equations \eqref{eq:Hjjbkmkn} and \eqref{eq:Pjjbkmkn} yield
to
\begin{align}
H_{n} & =\frac{1}{4}\sum_{m}k_{(+)n-m}\cdot k_{(+)m}+\frac{1}{4}\sum_{m}k_{(-)n-m}\cdot k_{(-)m}\;,\\
P_{n} & =\frac{1}{2}\sum_{m}k_{(+)n-m}\cdot k_{(-)m}\;.
\end{align}
In this basis, the Abelian current algebra acquires the form
\begin{align}
\left[k_{(+)m}^{I},k_{(-)n}^{J}\right] & =2mg^{IJ}\delta_{m+n,0}\;\;,\;\;\left[k_{(\pm)m}^{I},k_{(\pm)n}^{J}\right]=0\;,\label{eq:kmn-Algebra}
\end{align}
and the conformal generators fulfill
\begin{align}
\left[P_{m},P_{n}\right] & =\left(m-n\right)P_{m+n}\;\;,\;\;\left[P_{m},H_{n}\right]=\left(m-n\right)H_{m+n}\;,\label{eq:P,P-P,H-Algebra}\\
\left[H_{m},H_{n}\right] & =\left(m-n\right)P_{m+n}\;,\label{eq:H,H-Algebra}
\end{align}
so that the remaining brackets
\begin{equation}
\left[P_{n},k_{(\pm)m}^{I}\right]=-mk_{(\pm)m+n}^{I}\;\;,\;\;\left[H_{n},k_{(\pm)m}^{I}\right]=-mk_{(\mp)m+n}^{I}\;,\label{eq:P,H,kmn-Algebra}
\end{equation}
complete the semidirect sum.

\subsection{Deformed electric- and magnetic-like generators}

Here we consider two inequivalent finite $J\cdot\bar{J}$ deformations
of the energy density $H=T+\bar{T}$, given by $H_{(+)}$ and $H_{(-)}$,
defined as
\begin{equation}
H_{(\pm)}=T+\bar{T}\pm2J\cdot\bar{J}\;.\label{eq:HmpTTb}
\end{equation}
Changing the basis according to \eqref{eq:H,P} and \eqref{eq:kpm},
the deformed Hamiltonians are then given by
\begin{equation}
H_{(\pm)}=k_{(\pm)}^{2}\;.\label{eq:Hmpkmkp}
\end{equation}
In modes, equations \eqref{eq:HmpTTb} and \eqref{eq:Hmpkmkp} then
reduce to
\begin{align}
H_{(\pm)n} & =L_{n}+\bar{L}_{-n}\pm\sum_{m}J_{n+m}\cdot\bar{J}_{m}=\frac{1}{2}\sum_{m}k_{(\pm)n-m}\cdot k_{(\pm)m}\;,\label{eq:Hmn}
\end{align}
Note that $H_{(+)}$ is given by the square of a vector, whereas $H_{(-)}$
corresponds to the square of a pseudovector. Hence, hereafter we refer
to $H_{(+)}$ as the electric-like deformation, and to $H_{(-)}$
as the magnetic-like one.

One can then show that both independent sets of generators
\begin{align}
\text{Electric-like:} & \;\;\{H_{(+)},P,k_{(+)}^{I},k_{(-)}^{I}\}\;,\;\text{and}\label{eq: set-Electric}\\
\text{Magnetic-like:} & \;\;\{H_{(-)},P,k_{(+)}^{I},k_{(-)}^{I}\}\;,\label{eq:set-Magnetic}
\end{align}
fulfill a semidirect sum of the BMS$_{3}$ algebra with the current
algebra \eqref{eq:kmn-Algebra}. For both cases the brackets explicitly
read
\begin{align}
\left[P_{m},P_{n}\right] & =\left(m-n\right)P_{m+n}\;\;,\;\;\left[P_{m},H_{(\pm)n}\right]=\left(m-n\right)H_{(\pm)m+n}\;,\label{eq:PP-P,P,Hmn}
\end{align}
\begin{equation}
\left[H_{(\pm)m},H_{(\pm)n}\right]=0\;,\label{eq:Hmn,Hmn=00003D0}
\end{equation}
which span the BMS$_{3}$ subalgebra, while
\begin{align}
\left[P,k_{(\pm)m}^{I}\right] & =-mk_{(\pm)m+n}^{I}\;\;,\;\;\left[H_{(\pm)n},k_{(\mp)m}^{I}\right]=-2mk_{(\pm)m+n}^{I}\;,\label{eq:PK-BMS3}
\end{align}
\begin{equation}
\left[H_{(\pm)n},k_{(\pm)m}^{I}\right]=0\;,\label{eq:HmnKmn}
\end{equation}
complete the semidirect sum with the algebra of the currents in \eqref{eq:kmn-Algebra}.

It is worth highlighting that both finite electric- and magnetic-like
deformations yield to BMS$_{3}$ subalgebras, which correspond to
ultra/non-relativistic versions of the conformal algebra BMS$_{3}$$\thickapprox$
GCA$_{2}$$\thickapprox$ CCA$_{2}$, without the need of any kind
of limiting process.

One is then led to conclude that generic bosonic toroidal CFT$_{2}$
of the class under discussion admit finite electric- and magnetic-like
deformations describing BMS$_{3}$ field theories, being invariant
under conformal Carrollian symmetries. Indeed, expressing the action
of the CFT$_{2}$ in Hamiltonian form, which in the conformal gauge
reads
\begin{equation}
I_{CFT_{2}}=\int d^{2}x\,\left(\Pi\dot{\Phi}-H\right)\;,\label{eq:I_CFT}
\end{equation}
where $\Phi$ and $\Pi$ collectively stand for the fields and their
momenta, the corresponding electric- and magnetic-like finite deformations
are given by 
\begin{equation}
I_{BMS_{3}}^{(\pm)}=\int d^{2}x\,\left(\Pi\dot{\Phi}-H_{(\pm)}\right)\;,\label{eq:I_alpha-2}
\end{equation}
with $H_{(\pm)}$ defined through eq. \eqref{eq:HmpTTb}.

It is also worth stressing that since both electric- and magnetic-like
deformations of the CFT$_{2}$ preserve the algebra of the chiral
currents $J_{n}^{I}$ and $\bar{J}_{n}^{I}$, the quantization of
the deformed theories can be carried out in terms of the same (unitary)
representations of the original (undeformed) ones.

\section{Finite $J\bar{J}$ deformations of the bosonic string \label{sec:deformationsBosonicString-III}}

A concrete application of the results found in section \ref{sec:BMS-field-theories}
can be directly performed for the bosonic string, whose action can
be written as

\begin{equation}
I=-\frac{{\cal T}}{2}\int d^{2}\sigma\ \sqrt{-h}h^{\alpha\beta}\partial_{\alpha}X\cdot\partial_{\beta}X\;.\label{eq:Polyakov-I}
\end{equation}
It is then useful to express the action \eqref{eq:Polyakov-I} in
Hamiltonian form, being given by 
\begin{equation}
I=\int d^{2}\sigma\ \Big[\Pi\cdot\dot{X}-NH-N^{\sigma}P\Big]\;,\label{eq:HamiltonianAction-Polyakov}
\end{equation}
with $\Pi_{\mu}=\frac{\delta L}{\delta\dot{X}^{\mu}}$, and where
$N$, $N^{\sigma}$ stand for Lagrange multipliers (lapse and shift,
respectively). The constraints and the field equations then read
\begin{align}
H & =\frac{1}{2{\cal T}}\left(\Pi^{2}+{\cal T}^{2}X^{\prime2}\right)\;\;,\;\;P=\Pi\cdot X^{\prime}\;,\label{eq:ConstraintsBosoString}\\
\dot{X}^{\mu} & =\frac{N}{{\cal T}}\Pi^{\mu}+N^{\sigma}X^{\mu\prime}\;\;,\;\;\dot{\Pi}_{\mu}=\left({\cal T}NX_{\mu}^{\prime}+N^{\sigma}\Pi_{\mu}\right)^{\prime}\;.\label{eq:FieldEqsPolyakov}
\end{align}
Thus, the currents 
\begin{align}
J^{\mu} & =\frac{1}{2\sqrt{{\cal T}}}\left(\Pi^{\mu}-{\cal T}X^{\mu\prime}\right)\;\;,\;\;\bar{J}^{\mu}=\frac{1}{2\sqrt{{\cal T}}}\left(\Pi^{\mu}+{\cal T}X^{\mu\prime}\right)\;,\label{eq:CurrentsPolyakov}
\end{align}
satisfy the Abelian algebra in \eqref{eq:currentsPoissonB} in terms
of the inverse background metric $g^{\mu\nu}$, and they are conserved
on-shell, since by virtue of \eqref{eq:FieldEqsPolyakov} they fulfill
\begin{equation}
\dot{J}^{\mu}=\left[\left(N^{\sigma}-N\right)J^{\mu}\right]^{\prime}\;\;,\;\;\dot{\bar{J}}^{\mu}=\left[\left(N^{\sigma}+N\right)\bar{J}^{\mu}\right]^{\prime}\;.
\end{equation}
In order to apply the finite deformations, note that

\begin{equation}
J\cdot\bar{J}=\frac{1}{4{\cal T}}\left(\Pi^{2}-{\cal T}^{2}X^{\prime2}\right)\;,\label{eq:JJbarDef}
\end{equation}
and it is also useful to express the currents \eqref{eq:CurrentsPolyakov}
in the basis in \eqref{eq:kpm}, so that $k_{(\pm)}^{\mu}=\bar{J}^{\mu}\pm J^{\mu}$
explicitly read\textcolor{black}{
\begin{equation}
k_{(+)}^{\mu}=\frac{1}{\sqrt{{\cal T}}}\Pi^{\mu}\;\;,\;\;k_{(-)}^{\mu}=\sqrt{{\cal T}}X^{\mu\prime}\;.\label{eq:kmkn}
\end{equation}
}

\subsection{Electric-like deformation \label{sec:Electric-like deformation}}

The finite electric-like deformation of the bosonic string can be
readily implemented by deforming the Hamiltonian according to $H\rightarrow H_{(+)}$,
with $H_{(+)}$ defined as in eq. \eqref{eq:HmpTTb}. This amounts
to adding twice $J\cdot\bar{J}$ in \eqref{eq:JJbarDef} to the hamiltonian
$H$ in \eqref{eq:ConstraintsBosoString}; or equivalently, just writing
$H_{(+)}$ as in \eqref{eq:Hmpkmkp}, with $k_{(+)}^{\mu}$ given
by \eqref{eq:kmkn}, i.e.,\textcolor{black}{
\begin{align}
H_{(+)} & =H+2J\cdot\bar{J}=k_{(+)}^{2}=\frac{1}{{\cal T}}\Pi^{2}\;.\label{eq:Hm-Electric-String}
\end{align}
}Therefore, the electric-like deformation of the action reads
\begin{equation}
I_{(+)}=\int d^{2}\sigma\left[\Pi\cdot\dot{X}-N_{(+)}\Pi^{2}-N^{\sigma}\Pi\cdot X^{\prime}\right]\;,\label{eq:Im-Electric}
\end{equation}
with $N_{(+)}=N{\cal T}^{-1}$, so that the tension can be gauged
away by a suitable redefinition of the lapse function.

Note that the standard tensionless limit of the bosonic string (see
e.g., \cite{Gamboa:1990wh,Lindstrom:1990qb,Isberg:1992ia,Isberg:1993av,Bagchi:2013bga,Bagchi:2015nca})
is then precisely recovered from the finite electric-like deformation
(see also \cite{Bagchi:2022nvj}). Indeed, rescaling the lapse function
of the bosonic string action in \eqref{eq:HamiltonianAction-Polyakov}
according to $N=2\tilde{N}_{(+)}{\cal T}$, and then taking the tensionless
limit ${\cal T}\rightarrow0$, one finds $I_{(+)}$ in \eqref{eq:Im-Electric}
with $N_{(+)}\rightarrow\tilde{N}_{(+)}$.

It is also simple to verify that the action $I_{(+)}$ can be obtained
from a Carrollian limit, being similar to that of ``electric type''
in \cite{Henneaux:2021yzg}.

The Lagrangian version of the action $I_{(+)}$ can be explicitly
written by finding the momenta from its own field equation, which
read
\begin{equation}
\Pi^{\mu}=\frac{1}{2N_{(+)}}\left(\dot{X}^{\mu}-N^{\sigma}X^{\prime\mu}\right)\;,
\end{equation}
and replacing it back into $I_{(+)}$. Thus, $I_{(+)}=\int d^{2}\sigma\frac{1}{4N_{(+)}}\left(\dot{X}-N^{\sigma}X^{\prime}\right)^{2}$,
can be written in a manifestly covariant form as 
\begin{equation}
I_{(+)}=\int d^{2}\sigma\ \mathcal{\mathscr{V}}^{\alpha}\mathcal{\mathscr{V}}^{\beta}\partial_{\alpha}X\cdot\partial_{\beta}X\;,\label{eq:Im-electric covariant}
\end{equation}
where $\mathcal{\mathscr{V}}^{\alpha}=\frac{1}{2\sqrt{N_{(+)}}}\left(1,-N^{\sigma}\right)$
is a vector density of weight $\sfrac{1}{2}$. 

The invariance of the action $I_{(+)}$ under the BMS$_{3}$ algebra
can then be alternatively seen from diffeomorphisms that preserve
the vector density (${\cal L}_{\xi}\mathscr{V}^{\alpha}=0$), since
$\xi=\xi^{\alpha}\partial_{\alpha}$ close in the Lie bracket according
to BMS$_{3}$ algebra \cite{Isberg:1992ia,Isberg:1993av}.

\subsection{Magnetic-like deformation}

The finite magnetic-like deformation of the bosonic string is also
directly carried out by deforming the Hamiltonian in \eqref{eq:ConstraintsBosoString}
by subtracting twice $J\cdot\bar{J}$ in \eqref{eq:JJbarDef}. Thus,
$H\rightarrow H_{(-)}$, with $H_{(-)}$ given by \eqref{eq:HmpTTb},
or equivalently by \eqref{eq:Hmpkmkp}, with $k_{(-)}^{\mu}$ defined
by \eqref{eq:kmkn}, i.e.,\textcolor{black}{
\begin{align}
H_{(-)} & =H-2J\cdot\bar{J}=k_{(-)}^{2}={\cal T}X^{\prime2}\;.\label{eq:Hm-Magnetic-String}
\end{align}
}The magnetic-like deformation of the action is then given by
\begin{equation}
I_{(-)}=\int d^{2}\sigma\left[\Pi\cdot\dot{X}-N_{(-)}X^{\prime2}-N^{\sigma}\Pi\cdot X^{\prime}\right]\;,\label{eq:Im-Magentic}
\end{equation}
where $N_{(-)}=N{\cal T}$. 

The action $I_{(-)}$ thus yields to a new theory that is worth to
be explored. Note that the action \eqref{eq:Im-Magentic} is manifestly
covariant in the target space, so that the theory is still relativistic,
but endowed with an \textquotedblleft inner Carrollian structure\textquotedblright ,
being expressed by the BMS$_{3}$ algebra of the constraints. Furthermore,
the theory is clearly devoid of tension since it was gauged away by
a suitable redefinition of the lapse (differing from that in the electric-like
case).

One peculiarity of the magnetic-like deformed action $I_{(-)}$ is
that it is ``intrinsically Hamiltonian'', since the momenta $\Pi_{\mu}$
cannot be expressed in terms of $\dot{X}^{\mu}$ (nor $X^{\mu\prime}$). 

Interestingly, the classical solutions can be seen to describe a sort
of \textquotedblleft self-interacting null particle\textquotedblright{}
that moves along generic null curves of the original background metric,
not necessarily being geodesics. Indeed, the field equations are found
varying $I_{(-)}$ with respect to $X^{\mu}$ and $\Pi_{\mu}$, so
that
\begin{align}
\dot{X}^{\mu}-N^{\sigma}X^{\mu\prime} & =0\;,\label{eq:Xdot}\\
\dot{\Pi}^{\mu}-\left(2N_{(-)}X^{\mu\prime}+N^{\sigma}\Pi^{\mu}\right)^{\prime} & =0\;,\label{eq:Pidot}
\end{align}
while the constraints read
\begin{align}
H_{(-)} & =X^{\prime2}=0\;,\label{eq:H-magnetic}\\
P & =\Pi\cdot X^{\prime}=0\;.\label{eq:P-magnetic}
\end{align}
Choosing the gauge so that $N_{(-)}$ and $N^{\sigma}$ are constants,
the general solution of \eqref{eq:Xdot} is given by
\begin{align}
X^{\mu} & =X^{\mu}(\tilde{\sigma})\;,\label{eq:xsigma+}
\end{align}
where $\tilde{\sigma}=\sigma+N^{\sigma}\tau$. Thus, the Hamiltonian
constraint \eqref{eq:H-magnetic} implies that the curve described
by \eqref{eq:xsigma+} is null, so that $\tilde{\sigma}$ is an affine
parameter. The solution of the remaining field equation \eqref{eq:Pidot}
and the momentum constraint \eqref{eq:P-magnetic} then read
\begin{equation}
\Pi^{\mu}=Y^{\mu}+2N_{(-)}X^{\mu\prime\prime}\tau\;,\label{eq:Pi-mu-Magnetic}
\end{equation}
with $Y\cdot X^{\prime}=0$ and $Y^{\mu}=Y^{\mu}\left(\tilde{\sigma}\right)$. 

It is worth highlighting that the magnetic-like deformed action $I_{(-)}$
can also be attained from a nonstandard tensionless limit of the
bosonic string. In order to take the new limit, one first rescales
the fields and the momenta according to $X^{\mu}\rightarrow{\cal T}^{-1}X^{\mu}$,
$\Pi_{\mu}\rightarrow{\cal T}\Pi_{\mu}$, while the lapse is redefined
as $N=2{\cal T}\tilde{N}_{(-)}$. Hence, the action $I_{(-)}$ in
\eqref{eq:Im-Magentic}, with $N_{(-)}\rightarrow\tilde{N}_{(-)}$,
is recovered when ${\cal T}\rightarrow0$. It is also reassuring to
verify that not only the field equations \eqref{eq:Xdot}, \eqref{eq:Pidot},
and the constraints \eqref{eq:H-magnetic}, \eqref{eq:P-magnetic},
but actually the generic solution of the magnetic-like deformed action
in \eqref{eq:xsigma+}, \eqref{eq:Pi-mu-Magnetic} are also smoothly
obtained from the corresponding ones of the bosonic string in this
limit. Note that this nonstandard tensionless limit shares some similarity
with the Carrollian limit of ``magnetic type'' in \cite{Henneaux:2021yzg}
for the scalar field.

As an ending remark of this subsection, we point out that the magnetic-like
deformed action $I_{(-)}$ can also be recovered from alternative
limits in the tension. Indeed, one interesting possibility corresponds
to redefining the lapse according to $N=2{\cal T}^{-1}\hat{N}_{(-)}$,
without rescaling the fields and the momenta, so that the action $I_{(-)}$,
with $N_{(-)}\rightarrow\hat{N}_{(-)}$, is also recovered but now
with ${\cal T}\rightarrow\infty$. This possibility certainly looks
appealing since sending the string length to zero appears to go hand
in hand with the fact that the general solution of magnetic-like deformation
of the bosonic string describes a null curve instead of a surface. This
alternative limit also works well for the field equations and the
constraints; however, the generic solution of the magnetic-like deformed
action in \eqref{eq:xsigma+}, \eqref{eq:Pi-mu-Magnetic} is not smoothly
recovered in this case.

Further aspects of the magnetic-like deformation of the bosonic string
will be explored in a forthcoming work \cite{next}.

\section{BMS$_{3}$ deformations as limiting cases of continuous exactly marginal
(trivial) $J\bar{J}$ deformations \label{sec:BMS-deformationsIV}}

Here we show that the BMS$_{3}$ electric- and magnetic-like deformations
of the class of toroidal CFT$_{2}$'s discussed above, emerge from
limiting cases of certain continuous exactly marginal (trivial) current-current
deformations. Indeed, the absolute value of the current-current deformation
coupling turns out to be bounded, so that electric- and magnetic-like
BMS$_{3}$ field theories arise when the bound is saturated.

The class of deformations we are interested in corresponds to a $SO(1,1)$
subset of the full set of automorphisms of the algebra of left and
right currents in \eqref{eq:currentsPoissonB} that preserves the
momentum density, but spans a spectral flow in the energy density.
The $SO(1,1)$ automorphism of the currents is given by
\begin{align}
J_{(\alpha)}^{I} & =J^{I}\cosh\left(\frac{\alpha}{2}\right)+\bar{J}^{I}\sinh\left(\frac{\alpha}{2}\right)\;,\;\bar{J}_{(\alpha)}^{I}=\bar{J}^{I}\cosh\left(\frac{\alpha}{2}\right)+J^{I}\sinh\left(\frac{\alpha}{2}\right),\label{eq:J-alpha,Jbar-alpha}
\end{align}
so that $T_{(\alpha)}=J_{(\alpha)}^{2}$ and $\bar{T}_{(\alpha)}=\bar{J}_{(\alpha)}^{2}$
fulfill the conformal algebra. It is worth pointing out that
the $SO(1,1)$ automorphism induces a mixing of left and right sectors
given by
\begin{align}
T_{(\alpha)}= & T\cosh^{2}\left(\frac{\alpha}{2}\right)+\bar{T}\sinh^{2}\left(\frac{\alpha}{2}\right)+\sinh\left(\alpha\right)J\cdot\bar{J}\;,\nonumber \\
\bar{T}_{(\alpha)}= & \bar{T}\cosh^{2}\left(\frac{\alpha}{2}\right)+T\sinh^{2}\left(\frac{\alpha}{2}\right)+\sinh\left(\alpha\right)J\cdot\bar{J}\;,
\end{align}
but nevertheless, the momentum density $P_{(\alpha)}=\bar{T}_{(\alpha)}-T_{(\alpha)}$
does not change because it is manifestly invariant, since
\begin{equation}
P_{(\alpha)}=\bar{J}_{(\alpha)}^{2}-J_{(\alpha)}^{2}=P\;,
\end{equation}
while the energy density $H_{(\alpha)}=\bar{T}_{(\alpha)}+T_{(\alpha)}$
spectrally flows according to\footnote{Note that if the invariant bilinear form of the current algebra was
flat, i.e., $g_{IJ}=\delta_{IJ}$, the full set of automorphisms would
be given by $O(N,N)$, so that our $SO(1,1)$ automorphism in \eqref{eq:J-alpha,Jbar-alpha}
is a particular subset of the $\frac{O\left(N,N\right)}{O\left(N\right)\otimes O\left(N\right)}$
class of deformations of the energy density.}
\begin{equation}
H_{(\alpha)}=J_{(\alpha)}^{2}+\bar{J}_{(\alpha)}^{2}=\cosh\left(\alpha\right)H+2\sinh\left(\alpha\right)J\cdot\bar{J}\;.
\end{equation}
In the energy-momentum basis, the currents $k_{(\alpha)(\pm)}^{I}=\bar{J}_{(\alpha)}\pm J_{(\alpha)}$
transform as
\begin{align}
k_{(\alpha)(\pm)}^{I} & =e^{\pm\alpha/2}k_{(\pm)}^{I}\;.
\end{align}
Therefore, under the $SO(1,1)$ automorphism \eqref{eq:J-alpha,Jbar-alpha}
the original algebra does not change, i.e., the set $\{H_{(\alpha)},P,k_{(\alpha)(+)}^{I},k_{(\alpha)(-)}^{I}\}$
fulfills the same brackets as the original one $\{H,P,k_{(+)}^{I},k_{(-)}^{I}\}$
in eqs. \eqref{eq:kmn-Algebra}, \eqref{eq:P,P-P,H-Algebra}, \eqref{eq:H,H-Algebra}
and \eqref{eq:P,H,kmn-Algebra}, reflecting the fact that the automorphism
corresponds to exactly marginal (and actually trivial) deformations
for any given finite value of the parameter $\alpha$. Nonetheless,
in the limit $\alpha\rightarrow\pm\infty$, the corresponding deformations
are neither trivial nor marginal, because they actually describe BMS$_{3}$
electric- and magnetic-like deformations. In order to show that, it
is useful to rescale the energy density according to
\begin{equation}
\tilde{H}_{(\alpha)}=\frac{H_{(\alpha)}}{\cosh\left(\alpha\right)}=H+2\tanh\left(\alpha\right)J\cdot\bar{J}\;,\label{eq:Halphatilde}
\end{equation}
so that it can be equivalently written as
\begin{equation}
\tilde{H}_{(\alpha)}=\frac{1}{2}\left(1+\tanh\left(\alpha\right)\right)k_{(+)}^{2}+\frac{1}{2}\left(1-\tanh\left(\alpha\right)\right)k_{(-)}^{2}\;.\label{eq:Htildealphakmn}
\end{equation}
Thus, the semidirect sum of the conformal algebra with the currents
in eqs. \eqref{eq:kmn-Algebra}, \eqref{eq:P,P-P,H-Algebra}, \eqref{eq:H,H-Algebra},
\eqref{eq:P,H,kmn-Algebra}, now looks like
\begin{align}
\left[\tilde{\mathcal{H}}_{(\alpha)m},\tilde{\mathcal{H}}_{(\alpha)n}\right] & =\cosh^{-2}\left(\alpha\right)\left(m-n\right)J_{m+n}\;,\\
\left[\tilde{\mathcal{H}}_{(\alpha)n},k_{(\pm)m}^{I}\right] & =-\left(1\mp\tanh\left(\alpha\right)\right)mk_{(\mp)m+n}^{I}\;,
\end{align}
while the remaining commutators do not change. Therefore, when $\alpha\rightarrow\infty$
or $\alpha\rightarrow-\infty$ one obtains the semidirect sum of the
BMS$_{3}$ algebra with the currents for the electric- and magnetic-like
cases in eqs. \eqref{eq:PP-P,P,Hmn}, \eqref{eq:Hmn,Hmn=00003D0},
\eqref{eq:PK-BMS3}, \eqref{eq:HmnKmn}, respectively. Indeed, the
electric- and magnetic-like deformations of the energy density in
\eqref{eq:HmpTTb} are precisely recovered, since in the limits eq.
\eqref{eq:Halphatilde} reduces to 
\begin{equation}
\tilde{\mathcal{H}}_{(\pm\infty)}=H_{(\pm)}=H\pm2J\cdot\bar{J}\;,
\end{equation}
or equivalently, eq. \eqref{eq:Htildealphakmn} yields to \eqref{eq:Hmpkmkp}.

\section{Ending remarks \label{sec:EndingRemarks-V}}

The continuous deformations spanned by the $SO(1,1)$ automorphism
in \eqref{eq:J-alpha,Jbar-alpha} are of the form
\begin{equation}
H_{(g)}=H+g\;J\cdot\bar{J}\;,\label{eq:H(g)}
\end{equation}
with $g=2\tanh\left(\alpha\right)$, and hence the coupling is bounded
as $\left|g\right|\leq2$. These kind of deformations are exactly marginal
(and trivial) when the bound is strictly fulfilled ($\left|g\right|<2$),
and they become nontrivial when the bound is saturated, so that BMS$_{3}$
electric- and magnetic-like deformations correspond to $g=2$ and
$g=-2$, respectively.

One might naturally wonder about what occurs beyond the bound in the
coupling, i.e., for $\left|g\right|>2$, which certainly cannot be
attained from the $SO(1,1)$ automorphism in \eqref{eq:J-alpha,Jbar-alpha}.
In this case, it is useful to perform a rescaling of the form $\tilde{\mathcal{H}}_{(g)}=\left(\frac{g^{2}}{4}-1\right)^{-1/2}H_{(g)}$,
so that the sign on the right hand side of
\begin{equation}
\left[\tilde{\mathcal{H}}_{(g)m},\tilde{\mathcal{H}}_{(g)n}\right]=-\left(m-n\right)P_{m+n}\;,
\end{equation}
clearly yields to the Euclidean version of the conformal algebra (endowed
with the currents), c.f., \eqref{eq:H,H-Algebra}. However, note that
the deformations in this case do not describe a thermal version of
the original (undeformed) CFT$_{2}$, because the procedure does not
implement the corresponding Wick rotation. In other words, for a generic
gauge choice, the deformed Hamiltonian action for $\left|g\right|>2$
acquires the form

\begin{equation}
I_{(g)}=\int d^{2}x\ \Big[\Pi\dot{\Phi}-NH_{(g)}-N^{x}P\Big]\;,\label{eq:I(g)}
\end{equation}
where $H_{(g)}$ and $P$ span the (rescaled) Euclidean conformal
algebra, which misses the necessary overall imaginary unit ``$i$''
strictly required to describe the thermal version of the theory.

It is instructive to see how the current-current deformation works
for the whole range of the coupling $g$ in the case of the Lagrangian
(Polyakov) bosonic string \eqref{eq:Polyakov-I}. The deformed Hamiltonian
action becomes like \eqref{eq:I(g)}, where $H_{(g)}$ reads as in
\eqref{eq:H(g)}, while $H$, $P$ and $J\cdot\bar{J}$ are given
by \eqref{eq:ConstraintsBosoString} and \eqref{eq:JJbarDef}, so
that
\begin{eqnarray}
I_{(g)} & = & \int d^{2}\sigma\ \Big[\Pi\cdot\dot{X}-\frac{N}{2{\cal T}}\left\{ \left(1+\frac{g}{2}\right)\Pi^{2}+\left(1-\frac{g}{2}\right){\cal T}^{2}X^{\prime2}\right\} -N^{\sigma}\Pi\cdot X^{\prime}\Big]\;.\label{eq:I(g)Hamilt}
\end{eqnarray}
The momenta can be obtained from their own field equation, 
\begin{align}
\Pi^{\mu} & =\frac{{\cal T}}{N}\left(1+\frac{g}{2}\right)^{-1}\left(\dot{X}^{\mu}-N^{\sigma}X^{\prime\mu}\right)\;,\label{eq:Pi(g)}
\end{align}
and hence, the deformed Lagrangian action is recovered once \eqref{eq:Pi(g)}
is replaced back into \eqref{eq:I(g)Hamilt}, being given by
\begin{equation}
I_{(g)}=-\frac{{\cal T}}{2}\left(1+\frac{g}{2}\right)^{-1}\int d^{2}\sigma\ N^{-1}\left[-\dot{X}^{2}+2N^{\sigma}\dot{X}\cdot X^{\prime}-\left\{ \left(N^{\sigma}\right)^{2}-\left(1-\frac{g^{2}}{4}\right)N^{2}\right\} X'^{2}\Big]\right]\;.\label{eq:L(g)}
\end{equation}
It should be highlighted that this procedure does not apply in the
case of $g=-2$, since \eqref{eq:Pi(g)} becomes ill-defined. Therefore,
the magnetic-like deformation of the bosonic string cannot be directly
attained from the Lagrangian formalism, and so hereafter this case
is excluded.

The Lagrangian action $I_{(g)}$ can be written in a manifestly covariant
way with an inverse worldsheet metric that reads
\begin{equation}
h^{\alpha\beta}=\left(\begin{array}{cc}
-1 & N^{\sigma}\\
N^{\sigma} & -\left[\left(N^{\sigma}\right)^{2}-\left(1-\frac{g^{2}}{4}\right)N^{2}\right]
\end{array}\right)\;,
\end{equation}
so that the determinant of the worldsheet metric is given by
\begin{equation}
h=\left(\frac{g^{2}}{4}-1\right)^{-1}N^{-2}\;.
\end{equation}
We have then three cases:\medskip{}

(i) $\left|g\right|<2$: the worldsheet metric is Lorentzian, and
the Lagrangian action \eqref{eq:L(g)} just reduces to the Polyakov
action \eqref{eq:Polyakov-I}, but with an effective tension ${\cal T}\rightarrow{\cal T}_{(g)}={\cal T}\sqrt{\frac{2-g}{2+g}}$,
or equivalently, in terms of the parameter of the $SO(1,1)$ automorphism
${\cal T}\rightarrow{\cal T}_{(\alpha)}={\cal T}e^{-\alpha}$.\medskip{}

(ii) $g=2$: the worldsheet metric degenerates since $h=0$, and the
tension can be gauged away by a suitable rescaling of the lapse, so
that the electric-like deformation yields to the standard tensionless
string in \eqref{eq:Im-electric covariant}.

\medskip{}

(iii) $\left|g\right|>2$: the worldsheet metric is Euclidean, and
the Lagrangian action \eqref{eq:L(g)} also reduces to a Polyakov-like
one as in \eqref{eq:Polyakov-I}, with an effective tension ${\cal T}\rightarrow\tilde{{\cal T}}_{(g)}={\cal T}\sqrt{\frac{g-2}{g+2}}$.
In this case, the Lagrangian action then corresponds to a surface
with an Euclidean worldsheet metric but embedded into a Lorentzian
target manifold.\medskip{}

As an ending remark, it is worth mentioning that according to a very recent result \cite{He:2023lvk} (see also \cite{He:2022zcf})
in the context of celestial holography \cite{He:2015zea,Pasterski:2016qvg}
(for reviews see \cite{Strominger:2017zoo,Raclariu:2021zjz,Pasterski:2021rjz,Pasterski:2021raf}),
a bound in the coupling of current-current deformations would correspond
to a bound in the IR cutoff of the all-loop corrected scalar QED amplitudes
in four dimensions. In a future direction it would be worth exploring
whether the bound in the coupling of the current-current deformations
keep holding beyond the class of bosonic toroidal CFT's considered
here.

\section*{Acknowledgments}
We thank Arjun Bagchi, Aritra Banerjee, Glenn Barnich, Geoffrey Comp\`{e}re,
St\'{e}phane Detournay, Jos\'{e} Edelstein, Oscar Fuentealba, Gaston Giribet, Andr\'{e}s Gomberoff, Hern\'{a}n Gonz\'{a}lez, Marc Henneaux,
Diego Hidalgo, Javier Matulich, Alfredo P\'{e}rez, Miguel Pino, Pablo Rodr\'{i}guez and Patricio Salgado-Rebolledo for
useful comments and discussions. RT thanks the organizers of the Solvay
Workshop on ``Progress on gravitational physics: 45 years of Belgian-Chilean
collaboration'', during April 2023 in Brussels, for the opportunity
of presenting this work in a wonderful atmosphere. DT and RT also
thank the Physique Th\'{e}orique et Math\'{e}matique group of the Universit\'{e}
Libre de Bruxelles and the International Solvay Institutes for the
kind hospitality. This research has been partially supported by ANID
FONDECYT grants N$^{\circ}$ 1211226, 1220910, 1221624 and 3210558.

\bibliographystyle{JHEP}
\bibliography{CFT_Deformations_Arxiv}

%\providecommand{\href}[2]{#2}\begingroup\raggedright\begin{thebibliography}{10}

%\end{thebibliography}\endgroup

\end{document}